\documentclass[12pt,preprint]{aastex}
\usepackage{multirow} 	
\begin{document}

\title{Hot and cold spots counts as probes of non-Gaussianity in
  the CMB} 
\author{Pravabati Chingangbam\altaffilmark{1}, Changbom Park\altaffilmark{2},
K. P. Yogendran\altaffilmark{3}, Rien van de Weygaert\altaffilmark{4}}

\affil{\altaffilmark{1}Indian Institute of Astrophysics, Koramangala II Block, Bangalore 560034, India}
\affil{\altaffilmark{2}Korea Institute for Advanced Study, 85 Hoegiro,
  Dongdaemun-gu, Seoul 130-722, Korea} 
\affil{\altaffilmark{3}Indian Institute for Science Education and Research, Mohali, India} 
\affil{\altaffilmark{4}Kapteyn Astronomical Institute, University of Groningen, P.O.Box 800, 9747 AV Groningen, The Netherlands.}

\email{email : prava@iiap.res.in, cbp@kias.re.kr, pattag@gmail.com, weygaert@astro.rug.nl}

\begin{abstract}
We introduce the numbers of hot and cold spots, $n_h$ and $n_c$, of excursion
sets of the CMB temperature anisotropy maps as statistical
observables that can discriminate different non-Gaussian models. 
We numerically compute them from
simulations of non-Gaussian CMB temperature fluctuation maps. The first kind of
non-Gaussian model we study is the local type  primordial
non-Gaussianity. The second kind of models have some specific form  of
the probability distribution function from which the temperature
fluctuation value at each pixel is drawn, obtained using HEALPIX. We find the
characteristic non-Gaussian deviation shapes of $n_h$ and $n_c$, which
is distinct for each of the models under consideration. We further demonstrate that $n_h$ and $n_c$ carry additional information compared to the genus, which is just their linear combination, making them valuable additions to the Minkowski Functionals in constraining
non-Gaussianity.  
\end{abstract}


\maketitle

\section{Introduction} 

The statistical nature of the cosmic microwave background (CMB)
radiation temperature fluctuation field that 
we see today must be predominantly inherited from those of the
primordial density fluctuations. Inflation is currently the forerunner
amongst possible mechanisms that could have produced the primordial density
perturbations. All models of inflation, in general,  
predict some amount of deviation of these fluctuations from a Gaussian
distribution, with the details of the deviations being model
dependent. The knowledge of these deviations, if observed, for  
example in the CMB, will thus be of
much value in  distinguishing between various models of inflation. The
observational search for primordial non-Gaussianity, however, is not easy since
various observational effects can mask the true CMB signal. 
Given this difficulty the need for efficient, sensitive and complementary observables that can characterize non-Gaussian deviations cannot be  over
emphasized.   

Popular statistical measures of non-Gaussianity that can be obtained
from the CMB fall under two broad categories. The first are
observables that are defined in harmonic space~\citep{2011ApJS..192...18K,2010PhRvD..81l3007S,2011MNRAS.417..488C, 2010MNRAS.404..895V} such as the bispectrum,trispectrum, wavelets, the spherical Mexican hat wavelet etc. The second
category consists of those that directly exploit the geometric and topological properties of the temperature 
fluctuation field.  A popular class of observables, the Minkowski Functionals (MFs)~\citep{1986PThPh..76..952T,1988MNRAS.234..509C,1990ApJ...352....1G,1998MNRAS.297..355S,1998NewA....3...75W}, have long been applied to constrain non-Gaussianity in
the CMB. Considerable progress has been made in understanding them analytically for weakly non-Gaussian random
fields~\citep{2003ApJ...584....1M,2006ApJ...653...11H, 2008MNRAS.389.1439H, 2009ApJS..180..330K,2009PhRvD..80h1301P,2010PhRvD..81h3505M,2012PhRvD..85b3011G}. Of the three MFs that can be defined for a 2-dimensional random field, the third one known as
the genus is a topological quantity which depend on the global
properties of the random 
field. It is given by the difference of the numbers of hot and cold
spots at any given temperature fluctuation field value. The genus and
other MFs have non-Gaussian deviation shapes which is characteristic
of the non-Gaussian model. The non-Gaussian deviation shape of each
observable tells us what field values are best probed by the
observable, and these are the values where the deviations are the largest. 

In this paper, we introduce the numbers of hot and cold spots as statistical
observables in their own right. Just like the genus they are topological quantities which depend only on the global properties of the temperature fluctuation field. They have been studied earlier by Coles and Barrow~\citep{1987MNRAS.228..407C} and approximate formula are known for Gaussian and some non-Gaussian random fields~\citep{1983mit..book.....V}. 
Our goal here is to determine  their sensitivity and non-Gaussian deviation shapes as signatures of various kinds of non-Gaussianities. We have computed them using numerical
methods from simulated  non-Gaussian CMB maps. We first compute them
for simulations containing the so called local type primordial
non-Gaussian model parametrized by $f_{NL}$~\citep{2003ApJ...597...57L} and
$g_{NL}$~\citep{2009JCAP...12..019C}. We also compute them using
non-Gaussian  simulations provided by
HEALPIX~\citep{2005MNRAS.357....1R,2005ApJ...622..759G}. We have obtained the
characteristic non-Gaussian deviations for these different types of
input primordial non-Gaussianity.

We do not intend to address all the issues of observational systematic effects in this first paper and focus on the theoretical understanding of their behaviour upon potential non-Gaussianity.
The expectation in introducing them is that we can get additional information about non-Gaussian fields by
using these two observables in addition to the genus and other
MFs. Since the genus is given by the difference between these two numbers, 
in taking their linear combination we are throwing away some
information. This expectation will be most justified if the number of
hot spots is independent of the number of cold spots.   
As we will see in section 3, this is not always the case and whether
they are independent or not depends on the non-Gaussian model. For
example, for the  local type primordial non-Gaussian model
parametrized by $f_{NL}$ and $g_{NL}$, we find that they are related
to each other in a specific way.  However, even for the models where
they are related, there is still additional gain of information
coming from the fact that they have non-Gaussian deviation shapes
which are quite distinct from those of the genus and the other MFs and
hence they best extract non-Gaussian deviations of the field at values
different from the MFs. Therefore, they carry inormation complementary to the MFs. We further calculate the uncertainties in the numbers of hot and cold spots, taking into account cosmic variance, instrumental noise and sample boundaries at our choice of smoothing scales. We demonstrate that there exists
additional information in the numbers of hot and cold spots
compared to the genus, as shown in section 4.

This paper is organized as follows: in section 2 we briefly describe 
excursion sets and hot and cold spots, followed by our method for numerically computing  them.  We then show the results for the numbers of hot and cold  spots computed from Gaussian CMB temperature fluctuation maps. In section 3
we present the non-Gaussian deviations of the numbers of hot and
cold spots for the different kinds of non-Gaussianity that we have
studied, first, the local type non-Gaussianity and secondly for the
input non-Gaussian   PDF models  obtained using HEALPIX. In section 4 we  analyze how observational effects such as beam profiles, instrument noise and incomplete sky coverage, affect the numbers of hot and cold spots by computing them from simulations to which these effects have been added. Further, we compare their sensitivity to non-Gaussianity with that of the genus and show that they can give more information than the genus.
We end with a summary and discussion of the implications of our results in section 5. 

\section{Hot and cold spots counts}

Let $f\equiv (T(\hat n) - T_0)/T_0$ denote the CMB temperature
anisotropy field, where, $T_0$ denotes the mean temperature. 
By rescaling $f$ by its rms value, $\sigma_0$, we can define
the {\em threshold temperature}, $\nu \equiv f/\sigma_0$. At each value
of $\nu$, if we consider the set of all pixels that have values equal to
or above $\nu$ we obtain what is usually referred to as an {\em excursion set}. This set consists of  many {\em connected regions} into which the temperature  field `manifold' has fractured, and {\em holes} within those regions due to the excluded  pixels. As is commonly done in the literature, we call each connected region a {\em hot spot} and each hole a {\em cold spot}. For an excursion set indexed by $\nu$, we  define:
\begin{itemize}
\item $n_h \ \equiv $ number of hot spots, and 
\item $n_c\ \equiv $ number of cold spots.  
\end{itemize}
As we change $\nu$, the excursion  sets behave as though they are a
one-parameter family of spaces parametrized by $\nu$, and their
properties  such as the numbers of the hot and cold spots, change systematically as a function of $\nu$.
\begin{figure}[h]
\begin{center}
\resizebox{2.in}{2.in}{\includegraphics{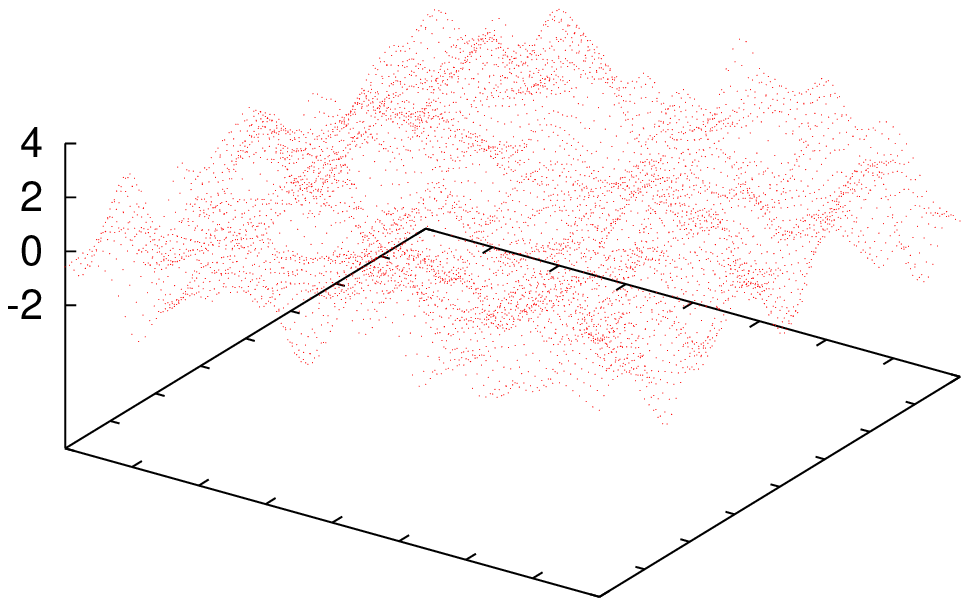}}
\resizebox{1.8in}{1.8in}{\includegraphics{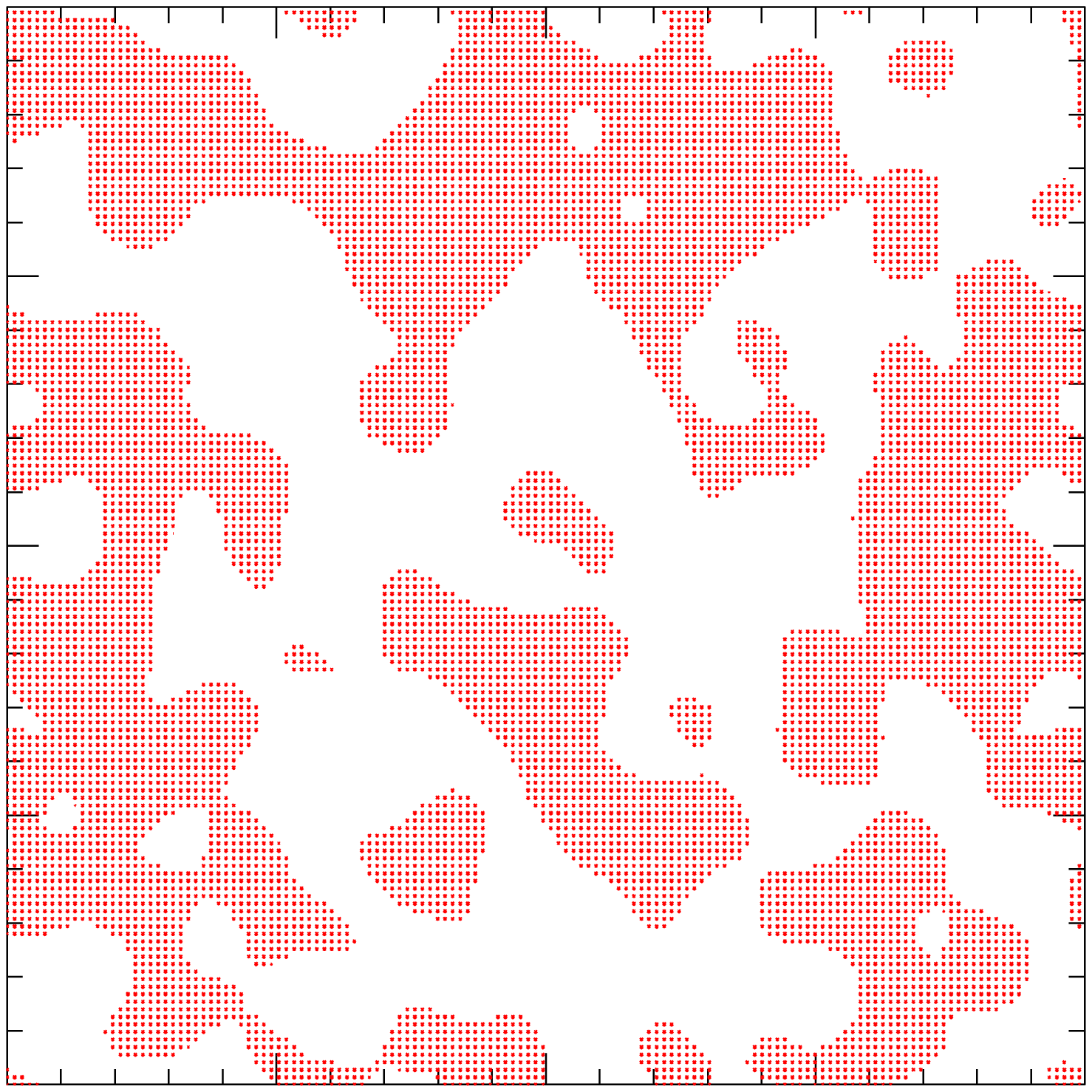}}
\resizebox{1.2in}{1.2in}{\includegraphics{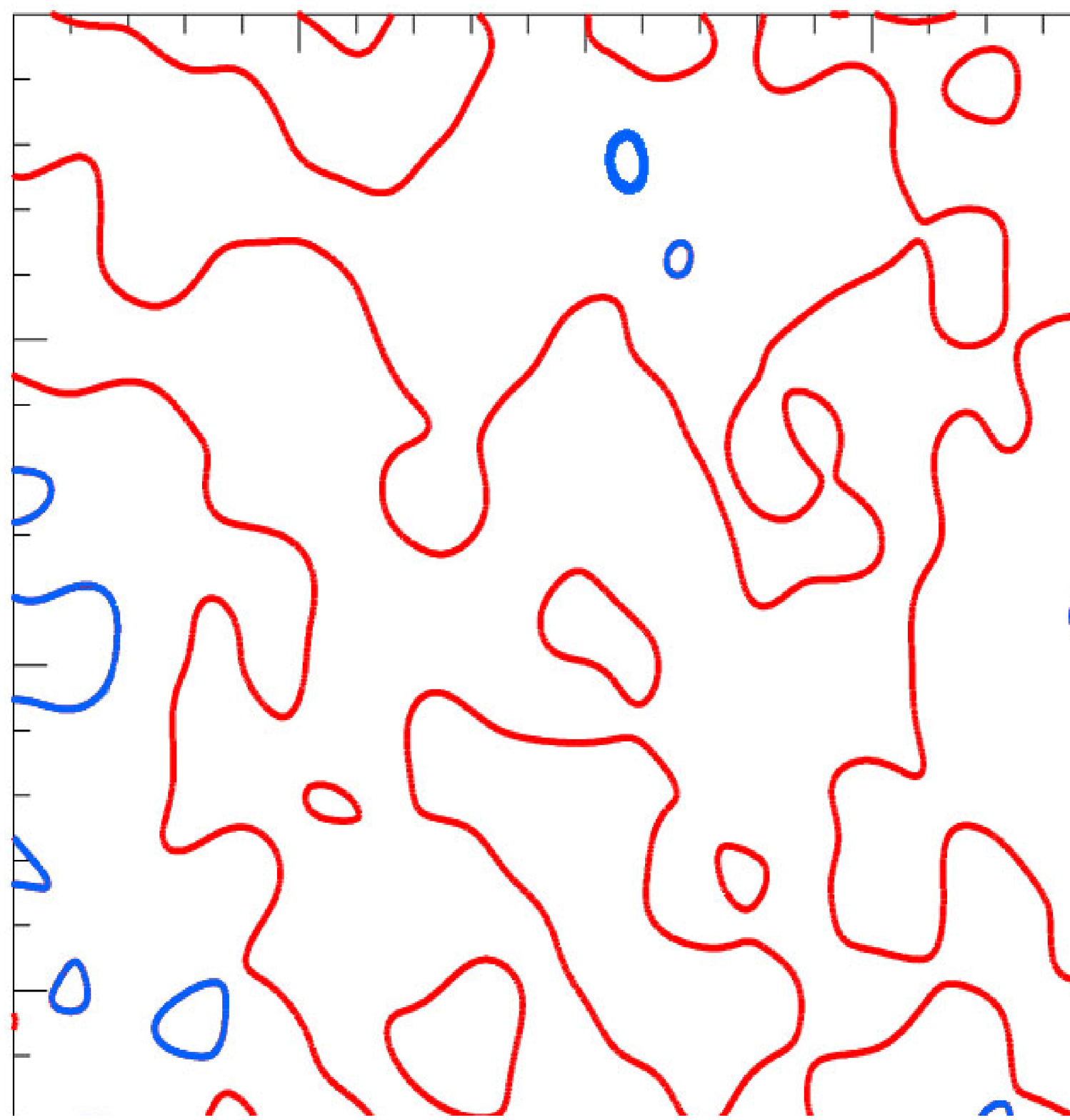}}
\end{center}
\caption{{\em Left panel}: A patch of a smoothed Gaussian fluctuation
  field. The field is defined on a square with periodic boundary condition. So
  field values that are at opposite ends of the square are
  identified. The $y$-axis gives the level value,
  $\nu$. {\em Middle panel}: The red regions gives the excursion set for   $\nu=0$. The  set is fragmented into several contiguous or connected regions. Some of the connected regions have holes within them. Each connected region is called a hot spot while each hole is called a cold spot.     
{\em Right panel}: Iso-temperature contours  enclosing
the excursion region (red lines) and holes (blue lines) for
$\nu =0$. One can see partial contours that are 
located at extreme ends of the square which together form closed contours. $n_h$ is the number of isolated connected regions, which can be obtained by
counting the closed red contours. $n_c$ is the number of  holes
within the connected regions and can be obtained by counting the blue
contours.} 
\label{fig:fieldcontour}
\end{figure}

We can relate the numbers of hot and cold spots to the numbers of
closed iso-temperature contours. The boundaries of each excursion set
are iso-temperature contours for the corresponding $\nu$. We can
assign an orientation to each of the contours - positive for those
that enclose hot spots and negative for the ones that enclose cold
spots. $n_h$ and  $n_c$ are then simply counts of closed positive and
negative orientation contours, respectively. For the purpose of
illustration,  in Fig.~(\ref{fig:fieldcontour}), we have shown a
smoothed Gaussian random field defined on a square with periodic
boundary condition. The left panel shows the full field. The middle
panel shows the excursion set for the same field for $\nu=0$. Connected regions and holes are clearly visible. The right
panel shows the boundary or iso-temperature contours for the same excursion
set, red enclosing hot spots and blue enclosing cold spots. 

Mathematically, we can express $n_h$ and  $n_c$ as line integrals
\begin{equation}
 n_h = \frac{1}{2\pi}\int_{C_+} K \,ds, \quad n_c = \frac{1}{2\pi}\int_{C_-} K \,ds,
\label{eqn:bettiformula}
\end{equation}
where $K$ is the total curvature of iso-temperature contours for each $\nu$. $C_+$ denotes contours that enclose hot spots while $C_-$ denotes contours that enclose cold spots. The genus, $g$, is given by a linear combination of $n_h$ and $n_c$: 
\begin{equation}
 g(\nu) \ = \ n_h(\nu)  - n_c(\nu).
\end{equation} 
For a Gaussian fluctuation field, $g$ is given by the expression 
\begin{equation}
 g(\nu) = A \,\nu\,\,e^{-\nu^2/2},
\end{equation}
where the amplitude $A$ carries the physical information about the
field and is given by 
$A = (1/2\pi^3) \left(\sigma_1/\sigma_0\right)^2$, 
with $\sigma_1$ being the rms of the gradient field. 

Note that we could equivalently define our excursion set at each $\nu$ to consist of the pixels having values below $\nu$, in which case the definitions of  $n_h$ and  $n_c$ would get interchanged.

\subsection{Numerical method for computing the numbers of hot and cold spots}
\label{sec:numerical}
Our method for computing the numbers of hot and cold spots is based on
the method for computing the genus  outlined in~\citep{1990ApJ...352....1G}. It 
is essentially an implementation of Eq.~(\ref{eqn:bettiformula}) based on connecting iso-temperature  pixels with the information of
the orientation of the contour retained.  At the end contours with
the same orientation are counted to get $n_h$ and  $n_c$.  

\begin{figure}[h]
\begin{center}
\resizebox{3.2in}{2.7in}{\includegraphics{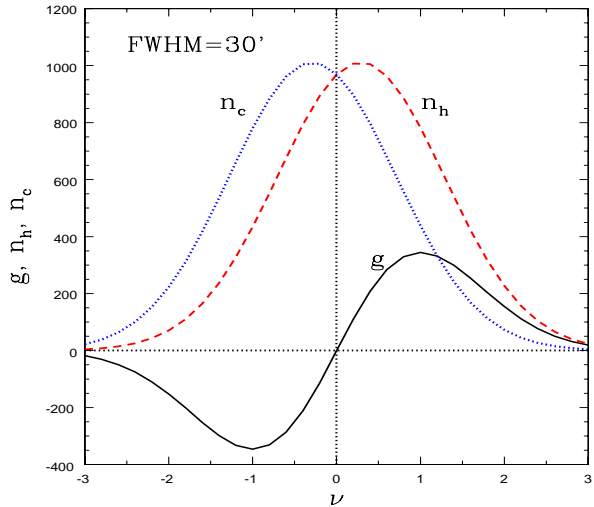}}
\end{center}
\caption{$n_h$, $n_c$ and $g$ measured from Gaussian simulations smoothed with FWHM=$30'$. The $y$ axis values are per unit area of the  sphere. The plots are average over 200 simulations.} 
\label{fig:genusnbetti}
\end{figure}

In Fig.~(\ref{fig:genusnbetti}) we have shown $n_h$, $n_c$ and $g$ versus $\nu$ obtained by averaging over measurements from 200 simulated Gaussian CMB anisotropy maps with the HEALPIX resolution parameter $Nside=512$ and smoothed with a Gaussian filter with FWHM$=30'$. The simulations have $\Lambda$CDM parameter values given by WMAP 5 years data~\citep{2009ApJS..180..330K}. The Gaussian genus formula serves to test the accuracy of the numerical computation of $n_h$ and $n_c$ (unless of course there is some error which contributes equally to both $n_h$ and $n_c$ and cancels out for the genus). 
It has been shown in~\citep{Changbom} that precise details of the
numbers of hot and cold spots in terms of Betti numbers for Gaussian random fields such
as the amplitude and location of peaks vary 
significantly as we vary the power index, $n$, of the input
three-dimensional power spectrum $P(k) \propto  k^{-n}$. The trend is that as $n$ increases the amplitude increases and the peak shifts closer to $\nu=0$. For $n \sim 3$, which is relevant for the CMB, the result that we have obtained is in agreement with this trend.


\section{Non-Gaussian deviations of the numbers of hot and cold spots}

In this section we compute the non-Gaussian deviations of numbers of hot and cold spots for different models of primordial non-Gaussianity. 

\subsection{Local type primordial non-Gaussianity}

\begin{figure}[h]
\begin{center}
\resizebox{6in}{5.5in}{\includegraphics{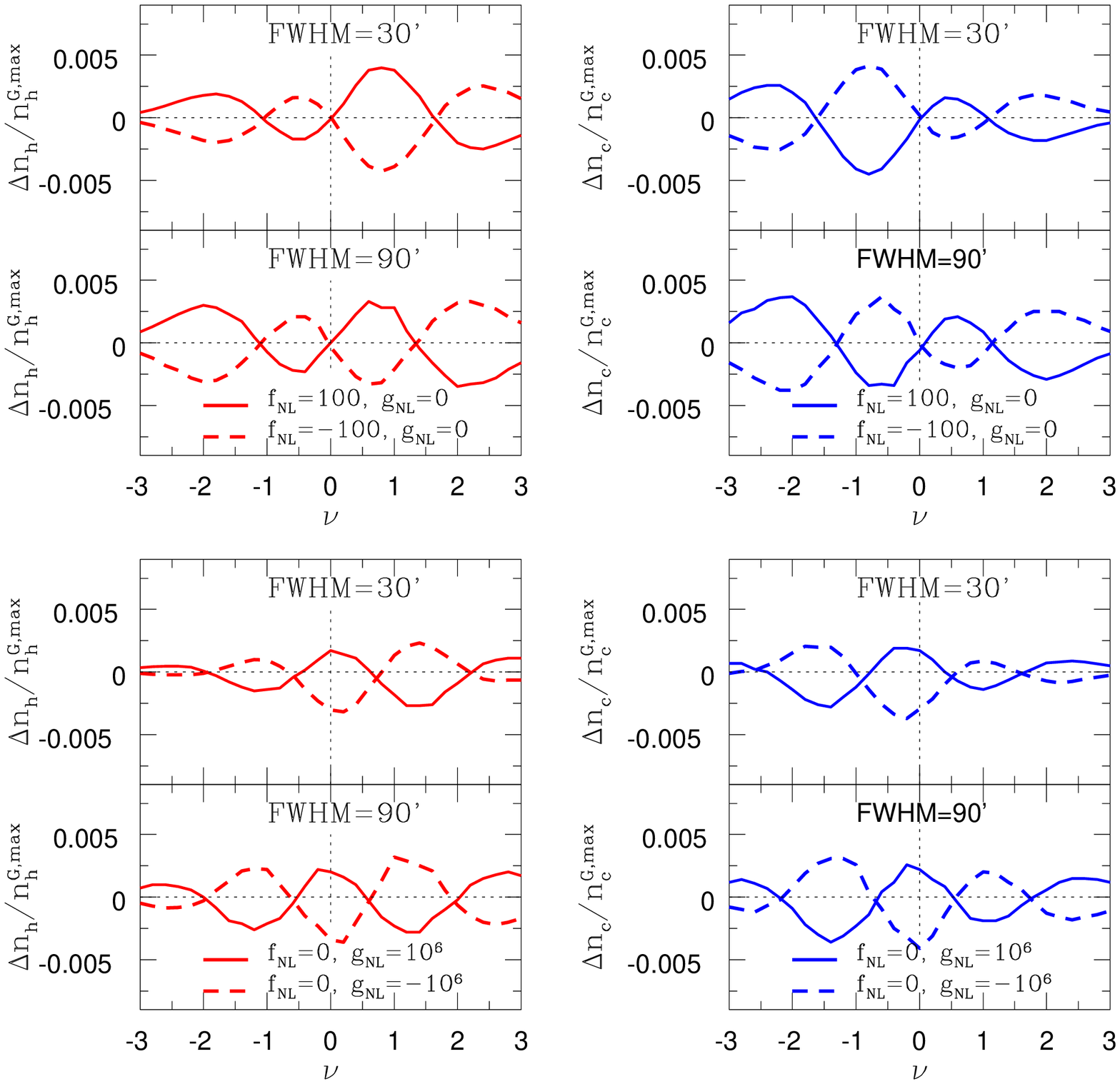}}
\end{center}
\caption{Non-Gaussian deviations of $n_h$ and  $n_c$ for pure $f_{NL}$
  (upper panels)  and pure $g_{NL}$ (lower panels)  input primordial
  non-Gaussianity at two smoothing angles - FWHM$=30'$ and
  $90'$. $\Delta n_i$ is defined as given in
  Eq.~(\ref{eqn:deltabeta}). $n_i^{G,max}$ is the maximum value
  of $n_i^G(\nu)$. 
  The $y$ axis values are per unit area of the sphere. The simulations have WMAP 5-years $\Lambda$CDM parameter values. The results are average over 200 simulations. } 
\label{fig:fnlgnlpurebetti}
\end{figure}

\begin{figure}[h]
\begin{center}
\resizebox{3.in}{2.5in}{\includegraphics{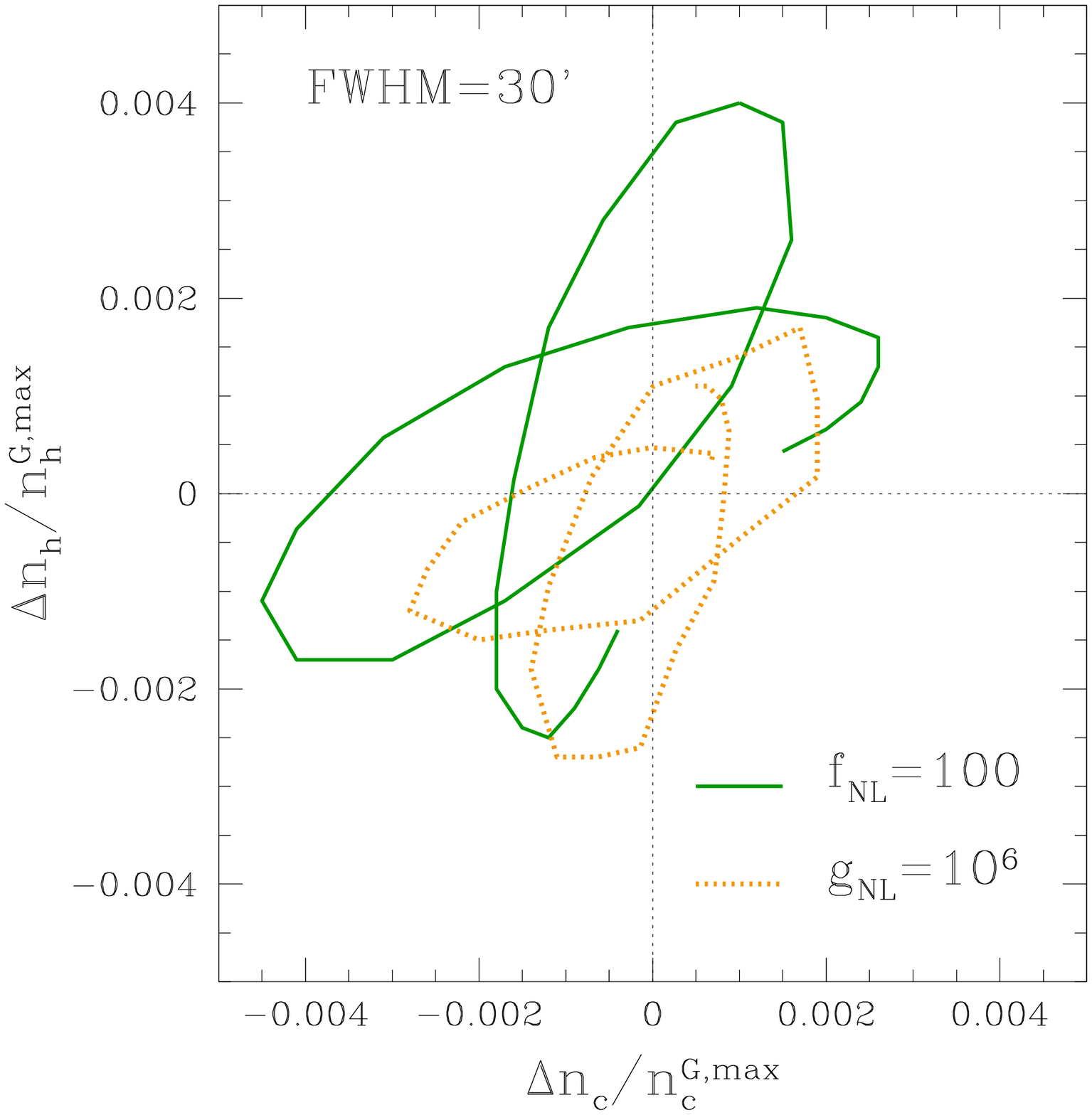}}
\end{center}
\caption{Plots of $\Delta n_h$ versus $\Delta n_c$ for pure $f_{NL}$
  (green, solid line) and for pure $g_{NL}$ (brown, dotted)
  models. This is a different way of showing the characteristics of
  the non-Gaussianity caused by the $f_{NL}$ and $g_{NL}$ terms.}
\label{fig:phasespace_fnlgnl}
\end{figure}
We consider the following expansion to cubic order of the primordial gravitational potential: 
\begin{equation}
\Phi(\vec x)= \Phi^G(\vec x) + f_{NL}\left( (\Phi^G(\vec
x))^2 - \langle   (\Phi^G)^2\rangle \right) +
g_{NL}(\Phi^G(\vec x))^3 + \ldots, 
\label{eqn:gravpotential}
\end{equation}
where $ \Phi^G$ is a Gaussian potential and $f_{NL}$ and $g_{NL}$ are constants which parametrize the first and second order non-linearities respectively, in the gravitational potential. Then, expanding the CMB temperature fluctuation field in multipoles, as $f = \sum_{\ell m} a_{\ell m}Y_{\ell m}$, we can calculate $a_{\ell m}$ by convolving $\Phi$ with the full radiation transfer function $\Delta_{\ell}$, as 
\begin{equation}
a_{\ell m}  =  4\pi (-i)^{\ell}\int \ \frac{d^3 k}{(2\pi)^3} \ \Phi({\vec k},t_i)   \    \Delta_{\ell}(k,t_0) \ Y^*_{\ell m}(\hat{k})
\end{equation}
We use simulations of non-Gaussian CMB maps~\citep{2003ApJ...597...57L,2009JCAP...12..019C} which have  Eq.~(\ref{eqn:gravpotential}) as the input potential to calculate $a_{\ell m}$. The input  power spectrum of $ \Phi^G$ is given as $P_{\Phi}(k)=(A_0/k^3) (k/k_0)^{n_s-1}$, where $A$, $n_s$ and $k_0$ are taken from WMAP 5 years $\Lambda$CDM parameter values~\citep{2009ApJS..180..330K}. The simulation resolution used is given by $NSIDE=512$, as in section~(\ref{sec:numerical}). We use $\Delta_{\ell}$ calculated with all perturbation terms kept to linear order~\citep{1996ApJ...469..437S} and hence the non-Gaussianity that shows up in the resulting CMB temperature field is a direct consequence of the primordial input. We have computed $n_h$ and $n_c$ for three kinds of simulations - pure $f_{NL}$, pure $g_{NL}$ and a mixture of the two. 
In order to quantify the non-Gaussian deviations we define 
\begin{equation}
\Delta n_i = n_i^{NG} - n_i^{G}, 
\label{eqn:deltabeta}
\end{equation}
where $i$ stands for $h$ or $c$, the index $G$  stands for Gaussian
and $NG$ for non-Gaussian. Plots are shown normalized by $n_i^{G,{\rm
    max}}$, which is the amplitude of $n_i^{G}$.  
 
\begin{figure}[h]
\begin{center}
\resizebox{6in}{5.5in}{\includegraphics{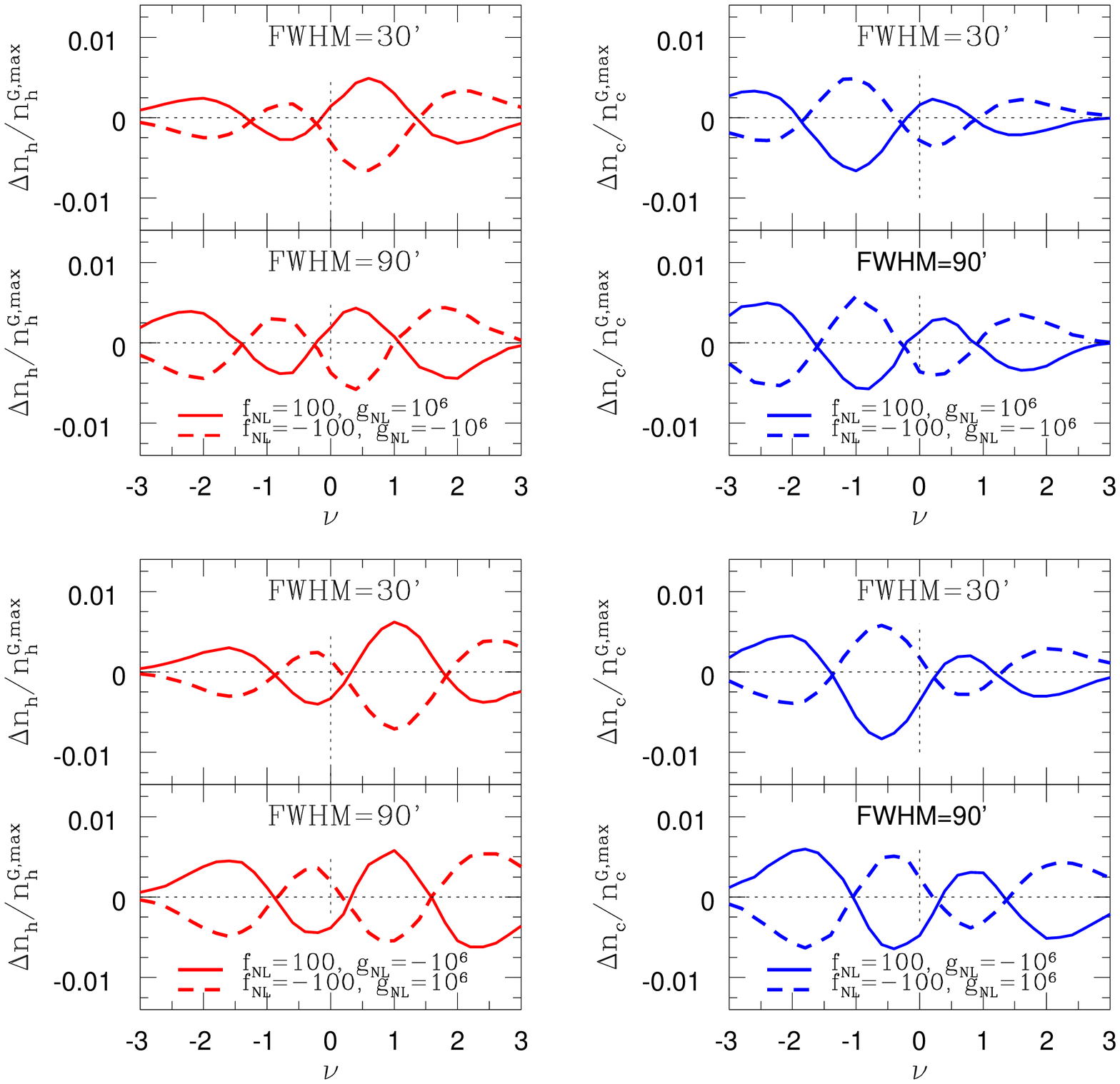}}
\end{center}
\caption{Same as in Fig.~(\ref{fig:fnlgnlpurebetti}) for mixture of  $f_{NL}$  and $g_{NL}$ input primordial non-Gaussianity.}
\label{fig:fnlgnlmixbetti}
\end{figure}

In Fig.~(\ref{fig:fnlgnlpurebetti}) we have plotted $\Delta n_h$ and
$\Delta n_c$  versus $\nu$ for pure $f_{NL}$ and pure $g_{NL}$
cases. We have used the values $f_{NL} = \pm 100$ and $g_{NL} = \pm
1\times 10^{6}$ and shown plots for two smoothing angles - FWHM$=30'$
and $90'$.   
For each case, it is important to note that $\Delta n_i$ has a characteristic non-Gaussian deviation shape and they can be easily distinguished from each other. There is slight variation of the deviation shapes as functions of the smoothing angle. 
Roughly speaking, the magnitude of the deviation at higher threshold values $|\nu| \gtrsim 2$ are larger for larger smoothing angles. 
An interesting observation is that for each case we can see that $n_h$ and $n_c$ are correlated as 
\begin{eqnarray}
\Delta n_h(\nu,f_{NL}) &=&  - \Delta n_c(-\nu,f_{NL}) \nonumber\\
\Delta n_h(\nu,g_{NL}) &=&  \Delta n_c(-\nu,g_{NL})
\label{eqn:spotscorrelate}
\end{eqnarray} 
In Fig.~(\ref{fig:phasespace_fnlgnl}) we have plotted $\Delta n_h$
versus $\Delta n_c$ at each value of $\nu$ for pure $f_{NL}$ (green,
solid line) and for pure $g_{NL}$ (brown, dotted) models. This is 
another way of visualizing the characteristics of the non-Gaussianity
caused by $f_{NL}$ or $g_{NL}$.
Fig.~(\ref{fig:fnlgnlmixbetti}) shows $\Delta n_i$ for the case when both  $f_{NL}$ and $g_{NL}$ contribute to the primordial non-Gaussianity for the same smoothing angles and parameter values. 
If we compare with non-Gaussian deviations of the genus (see Fig.(2)
of~\citep{2006ApJ...653...11H} for pure $f_{NL}$ case and Fig.(4)
of~\citep{2009JCAP...12..019C} for pure $g_{NL}$ case), we find that the
amplitude of the deviations of $n_h$ and $n_c$ are smaller by about
factor of two. 

We can get some idea about the dependence of $\Delta n_h$ and  $\Delta
n_c$  on $f_{NL}$ and $g_{NL}$ from the analytic expressions of the  non-Gaussian deviation of the genus~\citep{2006ApJ...653...11H,2003ApJ...584....1M}, which is given as an expansion in powers of $\sigma_0$. For pure $f_{NL}$ and pure $g_{NL}$ cases, keeping the genus expansion upto $\sigma_0$ and $\sigma_0^2$ orders, respectively, the genus non-Gaussian deviations have linear dependence on $f_{NL}$ and $g_{NL}$. 
When both $f_{NL}$ and $g_{NL}$ are present , then at $\sigma_0^2$ order there must be cross terms containing both $f_{NL}$ and $g_{NL}$.
Hence the non-Gaussian deviation of the genus will not be a simple linear combination of deviation terms depending on $f_{NL}$ and $g_{NL}$ independently. 
Since the genus is just the subtraction of  $n_c$ from   $n_h$, we can expect $n_c$ and $n_h$ to behave in a roughly similar fashion.


\subsection{HEALPIX non-Gaussian models}

We have generated non-Gaussian maps using the HEALPIX routine sky$\_$ng$\_$sim~\citep{2005MNRAS.357....1R,2005ApJ...622..759G}. This program implements two kinds of non-Gaussian models. The first is a model where the input probability distribution function is taken to be an expansion in excitepd states of the simple harmonic oscillator (SHO model), as given below: 
\begin{equation}
P(f) = e^{-f^2/2\sigma_0^2} \left| \sum_{i=0}^n \alpha_i C_i H_i\left(\frac{f}{\sqrt{2}\sigma_0}\right)   \right|^2,
\end{equation}
where $H_i$ are Hermite polynomials, $C_i$ are normalization constants, 
$\sigma_0$ is the variance of the Gaussian PDF and  $\alpha_i$, for $i\ge 1$,  are free parameters. $\alpha_0$ is constrained to be $\alpha_0 = \sqrt{1 - \sum_i^n|\alpha_i|^2}$. For our simulations we have kept terms upto $n=2$ such that $\alpha_1$ and $\alpha_2$ are non-zero. 
\begin{figure}[h]
\begin{center}
\resizebox{2.7in}{2.8in}{\includegraphics{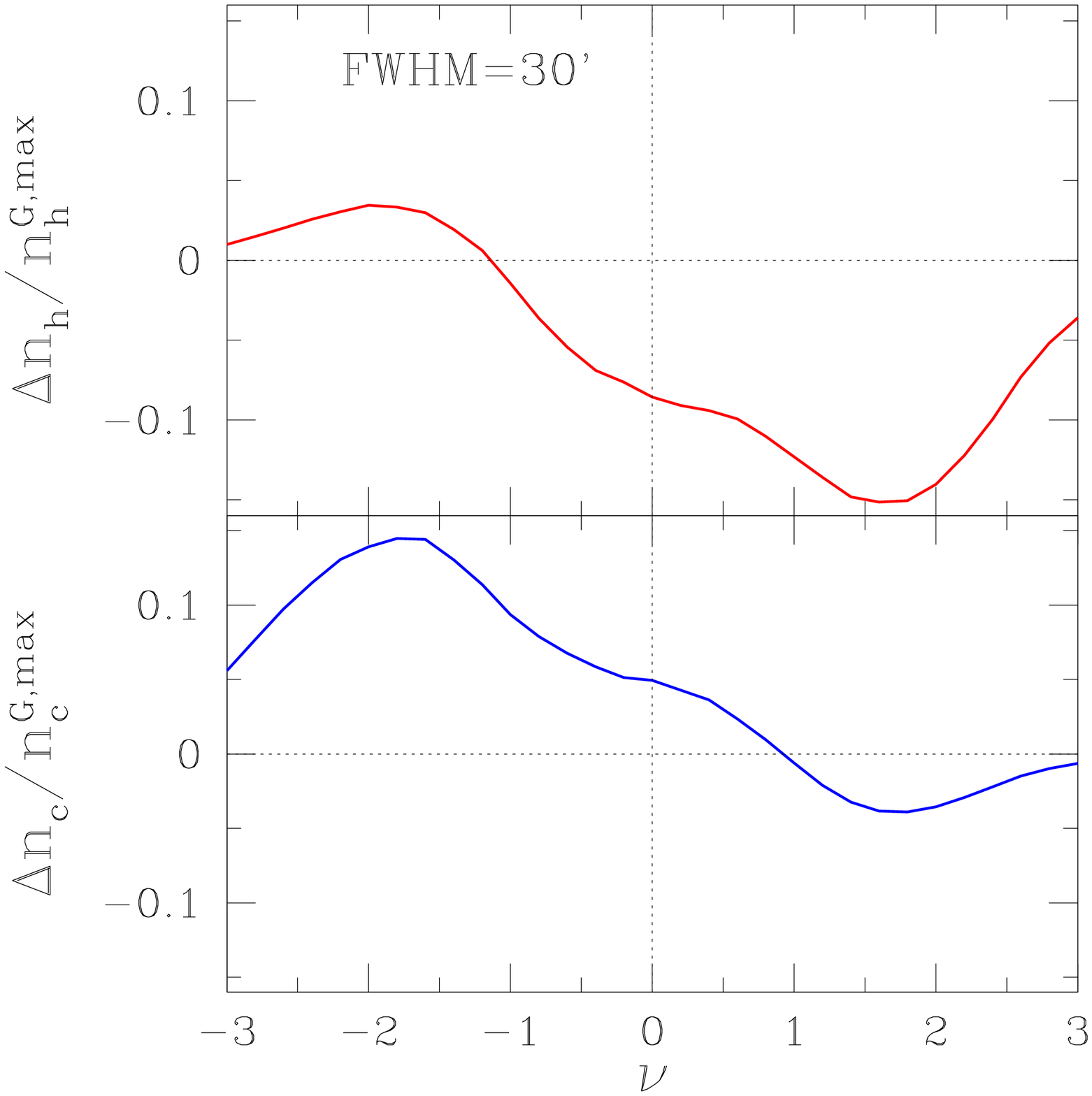}}
\quad
\resizebox{2.7in}{2.8in}{\includegraphics{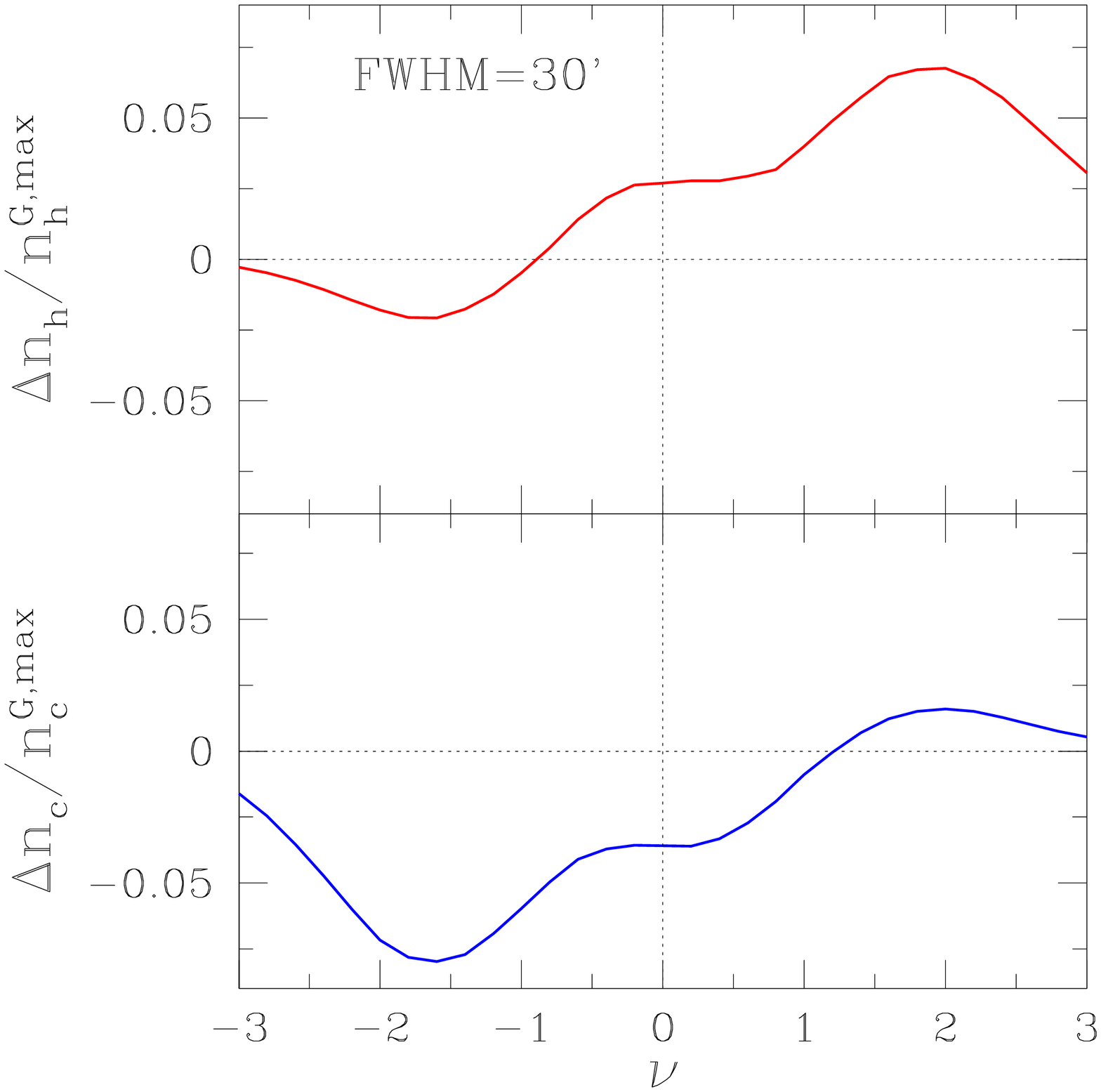}} 
\end{center}
\caption{{\em Left panel}: $\Delta n_i$ for the SH0 model. We have used $\alpha_1=0.6$ and $\alpha_2=0.6$.  {\em Right panel}:   $\Delta n_i$ for Gaussian power model. We have used $p=1$. The plots are average over 50 maps.}
\label{fig:sho_and_gausspower_betti}
\end{figure}
The second non-Gaussian model has the input PDF of the temperature field as an even power of a Gaussian PDF (Gaussian power model), with the temperature fluctuation value of the $k-$th pixel given by 
$$ f(k) = g^{2p}(k),$$ 
where $g$ is a zero mean, unit variance Gaussian variable, and $p$ is chosen to be a positive integer. We have used $p=1$ for our simulations. 

Fig.~(\ref{fig:sho_and_gausspower_betti}) shows the non-Gaussian
deviations of  $n_h$ and  $n_c$ for these two models. $\Delta n_i$ is
again defined as given in Eq.~(\ref{eqn:deltabeta}). The left panel
shows the deviations for SHO model and right panel shows for Gaussian
power model. As in the local non-Gaussianity case we can see
characteristic deviations for each type of non-Gaussianity. 

\begin{figure}[h]
\begin{center}
\resizebox{3in}{2.5in}{\includegraphics{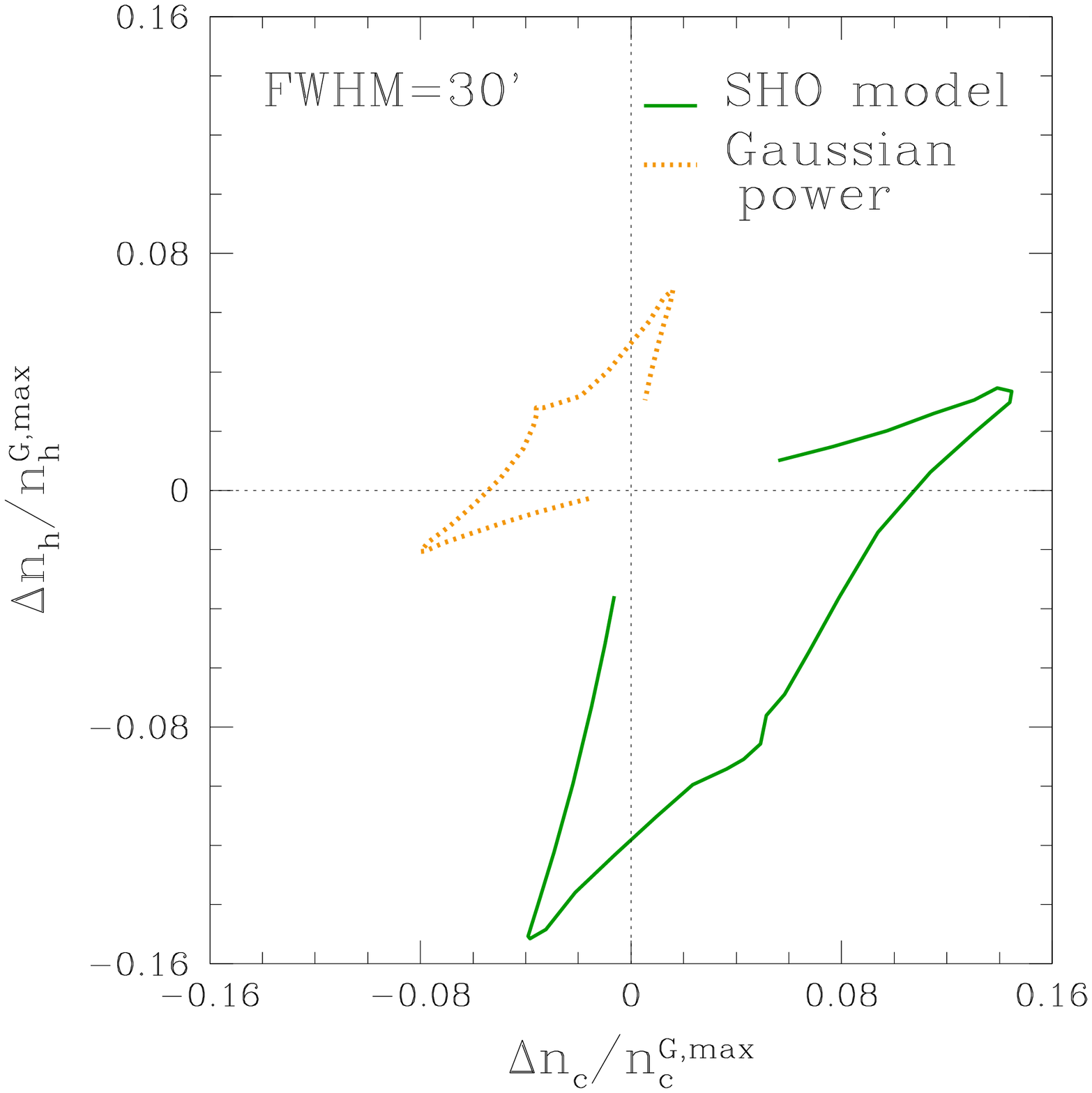}}
\end{center}
\caption{Plots of $\Delta n_h$ versus $\Delta n_c$ for SHO (green, solid line) and Gaussian power (brown, dotted line) models.}
\label{fig:phasespace_shogauss}
\end{figure}

For these models they do not have any simple correlation between
$\Delta n_h$ and  $\Delta n_c$ such as what we have seen for the local
type non-Gaussianity indicating that they carry information
independent from each other. Fig.~(\ref{fig:phasespace_shogauss})
shows $\Delta n_h$ versus $\Delta n_c$ at each value of $\nu$ for
these two models. These plots are a different way to characterize the
type of non-Gaussianity of the HEALPIX non-Gaussian models. 

\section{Statistical sensitivity of the numbers of hot and cold spots to $f_{NL}$ and $g_{NL}$}
To analyze the statistical power of $n_h$ and $n_c$ in realistic situations we measure them from simulations to which observational effects have been added.  
The observational effects are pixel window function, beam profile for each differential assembly (DA) and Gaussian noise realizations for each DA that follow the noise pattern, followed by Galaxy and point source masking. 
We then coadd $Q$, $V$ and $W$ DA's with appropriate weights obtained fron the inverse of the full-sky averaged pixel-noise variance in each DA, and then smooth the field. For the Galaxy masking we use the KQ75 mask.  
\begin{figure}[h]
\begin{center}
\resizebox{6in}{5.in}{\includegraphics{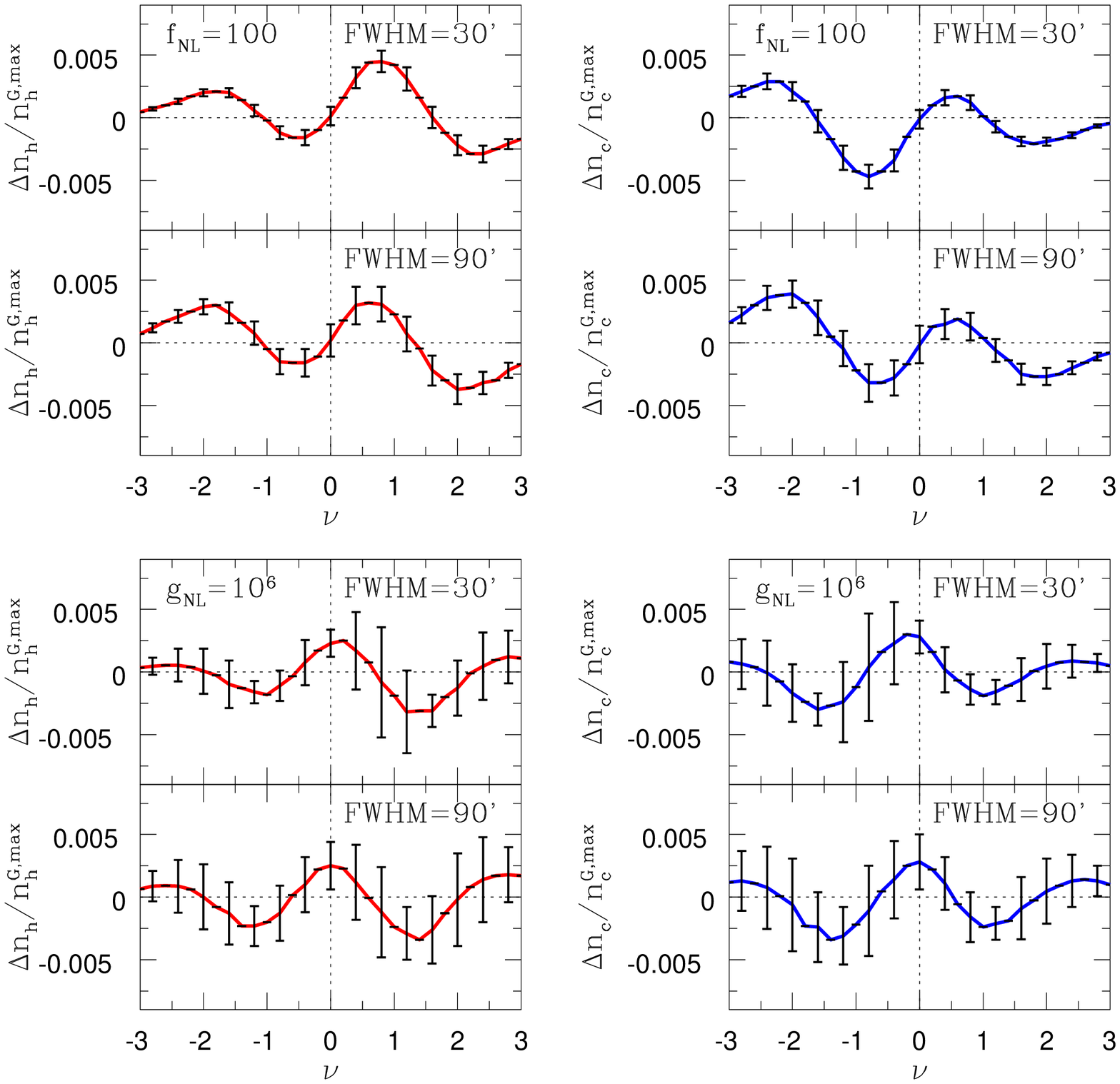}}
\end{center}
\caption{Non-Gaussian deviations of $n_h$ and  $n_c$ for $f_{NL}$
  (upper panels)  and $g_{NL}$ (lower panels) calculated after adding observational effects, namely pixel window function, beam profiles, noise for each DA and galaxy and point source masking, to the simulations. These calculations are from $Q+V+W$ coadded maps. 
The error bars are the sample variance from 200 simulations.}
\label{fig:fnlgnlbettionesigma}
\end{figure}
In Fig.~(\ref{fig:fnlgnlbettionesigma}) we have shown the sample variance error bars obtained from the 200 $Q+V+W$ coadded maps prepared as described above. It is immediately noticeable that the error bars for $g_{NL}$ is larger than those of $f_{NL}$ at each corresponding smoothing angle. This can be understood from Eq.~(\ref{eqn:gravpotential}) as follows. 
Suppose we have a perfectly Gaussian field $\phi^G$ 
and another `Gaussian' field with slight statistical fluctuations
$\phi' = \phi^G(1+D)$, where $D$ quantifies the fluctuation. Then the deviation from the Gaussian field when $f_{NL}$ and $g_{NL}$ contributions are present is given by $\Delta \phi^{NG} = D\phi^G + 2D\,f_{NL}(\phi^G)^2 + 3D\,g_{NL} (\phi^G)^3$. Hence statistical fluctuations seen for $g_{NL}$ will be larger than those for $f_{NL}$.

As a simple way of estimating the statistical disciminating power of the numbers of hot and cold spots in comparision to the genus we integrate the absolute values of the non-Gaussian deviations 
measured in units of the corresponding sample variances from $\nu=-3$ to 3. Let us denote it by $A$. For $M$ threshold levels with spacing $\Delta\nu$, we can calculate it as,
\begin{equation}
A=\Delta\nu\sum_{i=1}^M\left(|\Delta O(i)|/O^{G,max}\right)/\sigma_s(i),
\label{eqn:area}
\end{equation} 
where $O$ can be either $g$, $n_h$ or $n_c$, and $\sigma_s(i)$ are the respective sample variances at each threshold level $i$. For our case, $M=31$ and $\Delta\nu=0.2$. The resulting values are shown in Table (\ref{table:area}). We find considerably larger values of $A$ for $n_h$ and $n_c$ compared to $g$ both for $f_{NL}$ and $g_{NL}$ type non-Gaussianities, at the smoothing angles we have considered. This demonstrates that there is loss of statistical power for detecting the presence of non-Gaussian deviations when we combine $n_h$ and $n_c$ to get the genus.

\begin{center}
\begin{table}
\begin{tabular}{|c|c|c|r|}
\hline
Non-Gaussian input &  Smoothing FWHM & Observable & $A$ \\
\hline
\multirow{6}{1.9cm} {$f_{NL}=100$} & \multirow{3}{1.cm} {$30'$} &  g & 7.4\\ 
& &  $n_h$  &  11.3\\ 
& & $n_c$  &  11.3\\ \cline{2-4}
& \multirow{3}{1.cm} {$90'$} &  g & 3.3\\ 
& &  $n_h$  &  7.3\\ 
& & $n_c$  &  7.5\\
\hline
\multirow{6}{1.9cm} {$g_{NL}=10^6$} & \multirow{3}{1.cm} {$30'$} &  g & 2.0\\ 
& &  $n_h$  &  2.5\\ 
& & $n_c$  &  2.5\\ \cline{2-4}
& \multirow{3}{1.cm} {$90'$} &  g & 1.1\\ 
& &  $n_h$  &  2.2\\ 
& & $n_c$  &  2.1\\
\hline
\end{tabular}
\caption{Table showing values of $A$ defined in Eq.~(\ref{eqn:area}) for $g$, $n_h$ and $n_c$. }
\label{table:area}
\end{table}
\end{center}

\section{Conclusion}

We have introduced the numbers of hot and cold spots of the CMB temperature fluctuation field as statistical observables in their own right and propose to use them as discriminants of non-Gaussianity. We have studied the theoretical predictions for the numbers of hot and cold spots  and their expected non-Gaussian deviations for various kinds of non-Gaussianities. We have calculated them using numerical methods from simulated CMB maps containing the different non-Gaussian models as inputs.  
The first type of input non-Gaussian model we studied is the so called local type primordial non-Gaussianity, parametrized by  $f_{NL}$ and $g_{NL}$ at the first and second order non-linearity, respectively, of the perturbative expansion of the primordial gravitational potential. This gravitational potential is convolved with the full radiation transfer kept to linear order to obtain the simulated CMB temparature field, and hence the non-Gaussian deviations seen in the numbers of hot and cold spots are direct probes of the primordial non-Gaussianity. For these local primordial non-Gaussian models, what we find is that $n_h$ and $n_c$ are correlated as given by Eq.~(\ref{eqn:spotscorrelate}). The strengths of the non-Gaussian deviations of $n_h$, $n_c$ and $g$ are large at different regions of $\nu$ and hence each of them best probe regions of the field values specific to it. Therefore, they provide complementary information. Further, we have demonstrated that there exists additional information in the numbers of hot and cold spots
compared to their linear combination given by the genus. 

The second class of non-Gaussian models that we have considered assumes specific forms of the PDF of the temperature fluctuation field. In particular, we studied a model where the simulated temperature value at each pixel is drawn from a PDF given as an expansion in simple harmonic oscillator states. We also studied another model where the temperature fluctuation values at each pixel  is given as even powers of a number drawn from a Gaussian distributed field. Note that the assumption of the form of the PDF does not tell us anything about the physical source of the non-Gaussianity. 
Even though the physical origin of the non-Gaussianity is not clear, they are quite interesting models because they provide examples of non-Gaussian models where  $n_h$ and $n_c$ are not correlated. 

It is interesting to compare the shapes of the numbers of hot and cold spots  with those of maxima and minima counts~\citep{2011PhRvD..84h3510P,1986ApJ...304...15B,1987MNRAS.226..655B,1981grf..book.....A}. The shapes of $n_h$ and $n_c$ are roughly similar to the maxima and minima counts, respectively, though the precise shape information such as peak location and the amplitude is quite different. Note that the extrema counts will measure more number of objects per unit area in comparision to the numbers of hot and cold spots since a typical connected/hole region can have more than one maxima/minima. As $|\nu|$ becomes much larger than one, $n_h$ and $n_c$ should tend towards the maxima and minima counts, respectively. 

Our next goal is to apply the number of hot and cold spots to observational data and constrain $f_{NL}$ and $g_{NL}$. It would also be very useful to have  their analytic expressions. We are presently working towards these directions.

\acknowledgments
We thank Korea Institute for Advanced Study for providing computing resources (KIAS Center for Advanced Computation Linux Cluster System QUEST) where the local non-Gaussian simulations used in this paper were computed.   
We also acknowledge use of the Hydra cluster at the Indian Institute of Astrophysics for a part of the analysis.  
We acknowledge use of the HEALPIX package. 


\bibliographystyle{apj}
\bibliography{chingangbam.references}

\end{document}